\documentclass[prl,aps,showpacs,twocolumn,floatfix]{revtex4}
\usepackage{epsfig} \usepackage{graphics} \usepackage{bm}
\usepackage{amssymb}
\usepackage{graphicx}
\begin{document}

\preprint{Lebed-Rapids-LN}

\title{Layered Superconductor in a Magnetic Field: Breakdown
of the Effective Masses Model}

\author{A.G. Lebed$^*$}

\affiliation{Department of Physics, University of Arizona, 1118 E.
4-th Street, Tucson, AZ 85721, USA}

\begin{abstract}
We theoretically study the upper critical magnetic fields at zero
temperature in a quasi-two-dimensional (Q2D) superconductor in the
parallel and perpendicular fields, $H_{c2}^{\parallel}(0)$ and
$H_{c2}^{\perp}$(0), respectively. We find that
$H_{c2}^{\parallel}(0) \approx 0.75 \ | d H_{c2}^{\parallel}/ d
T|_{T_c} T_c $ and that $H_{c2}^{\perp}(0) \approx 0.59 \ | d
H_{c2}^{\perp}/ d T|_{T_c} T_c$, where $| d H_{c2}^{\parallel}/ d
T|_{T_c}$ and $| d H_{c2}^{\perp}/ d T|_{T_c}$ are the
corresponding Ginzburg-Landau slopes of the upper critical
magnetic fields. Our results demonstrate the breakdown of the
so-called effective mass model in Q2D case and may be partially
responsible for the experimentally observed deviations from the
effective mass model in a number of layered superconductors,
including $MgB_2$.
\end{abstract}


\maketitle

The upper critical magnetic field, $H_{c2}(T)$, is known to be one
of the most important properties of the type-II superconductors.
It destroys superconductivity due to the orbital Meissner currents
in case, where we can disregard the Pauli spin-splitting
paramagnetic effects. The Ginzburg-Landau (GL) theory gave tools
to calculate a slope of the $H_{c2}(T)$ [1] in the vicinity of
superconducting transition temperature, $(T_c-T)/T_c \ll 1$. On
the other hand, at zero temperature, the upper critical magnetic
field was calculated for an isotropic $3D$ superconductor in
Ref.[2]. Temperature dependence of $H_{c2}(T)$ in a whole
temperature region in an isotropic $3D$ superconductor was
calculated later in Ref.[3]. Important generalization of the GL
theory to the case of anisotropic superconductors was obtained in
Ref.[4], where the so-called effective mass model was implicitly
introduced. The effective mass model, partially based on the
results obtained in Ref. [4] in the GL region, states more: ratios
of the upper critical magnetic fields measured along fixed
different directions do not much depend on temperature. Recently
observed experimental temperature dependencies of anisotropy of
the upper critical fields in layered compound MB$_2$ [5] and other
materials are prescribed exclusively to many-band effects (see
introductory part of review [6]).

The goal of our Letter is to consider the orbital effect in a
parallel magnetic field in a Q2D conductor at zero temperature,
where we explicitly take into account a Q2D anisotropy of the
electron spectrum. In contrast to Refs.[1-4,6], we demonstrate
that, in a Q2D case in a parallel magnetic field, the solution of
the so-called gap equation can not be expresses as some
exponential function. Moreover, we show that the above mentioned
solution even changes a sign with changing space coordinate. This
leads to unusual value of the corresponding coefficient, $0.75$,
in the equation,
\begin{equation}
H_{c2}^{\parallel}(0) \approx 0.75 \ | d H_{c2}^{\parallel}/ d
T|_{T_c} T_c \ ,
\end{equation}
for a parallel magnetic field. We recall that, for a perpendicular
magnetic field the corresponding solution is exponential one and
gives much smaller coefficient - $0.59$ [7]:
\begin{equation}
H_{c2}^{\perp}(0) \approx 0.59 \ | d H_{c2}^{\perp}/ d T|_{T_c}
T_c \ .
\end{equation}
Note that Eqs.(1) and (2) directly break the effective mass model,
since the corresponding coefficients, $0.75$ and $0.59$ are not
close to each other. We also stress that, while deriving Eqs.(1)
and (2), we do not take into account quantum effects of electron
motion in a magnetic field [8-10].

In the Letter, we consider a layered superconductor with the
following realistic Q2D electron spectrum:
\begin{eqnarray}
&\epsilon({\bf p})= \frac{1}{2m} (p^2_x + p^2_y) - 2 t_{\perp}
\cos(p_z c^*) \ ,
\nonumber\\
&t_{\perp} \ll \epsilon_F \ , \ \ \ \ \epsilon_F =
\frac{p^2_F}{2m}=\frac{mv^2_F}{2} \ ,
\end{eqnarray}
where $m$ - the electron in-plane mass, $t_{\perp}$ - the integral
of overlapping of electron wave functions in a perpendicular to
the conducting planes direction; $\epsilon_F , \ p_F$, and $v_F$
are the Fermi energy, Fermi momentum, and Fermi velocity,
correspondingly; $\hbar \equiv 1$.  In a parallel to the
conducting planes magnetic field,
\begin{equation}
{\bf H} = (0,H,0) \  ,  \ \ \ \ {\bf A} = (0,0,-Hx) \ ,
\end{equation}
we make use of the so-called Peierls substitution method:
\begin{eqnarray}
&&p_x \rightarrow - i \biggl( \frac{\partial }{ \partial x}
\biggl), \ \ p_y \rightarrow - i \biggl( \frac{\partial }{
\partial y} \biggl),
\nonumber\\
&&c^* p_z \rightarrow - i c^* \biggl( \frac{\partial }{\partial z}
\biggl) - \biggl( \frac{\omega_c}{v_F} \biggl)x, \ \ \omega_c(H) =
\frac{e v_F c^* H}{c}.
\end{eqnarray}

Under such conditions the electron orbital Hamiltonian in a magnetic field
can be written in the following way:
\begin{eqnarray}
\hat H = - \frac{1}{2m} \biggl( \frac{\partial^2 }{ \partial x^2}
+ \frac{\partial^2 }{ \partial y^2} \biggl) - 2 t_{\perp} \cos
\biggl(-i c^* \frac{\partial}{\partial z} - \frac{\omega_c}{v_F}x
\biggl).
\end{eqnarray}
As directly follows from Eq.(6), electron wave functions can be
represented as
\begin{eqnarray}
&\Psi^{\pm}_{\epsilon}(x,y,z) = \exp[\pm i p^0_x(p_y) x] \ \exp( i
p_y y) \ \exp( i p_z z)
\nonumber\\
&\times \Phi_{\epsilon}^{\pm}(x, p_y,p_z), \ \  p^0_x(p_y) =
\sqrt{p_F^2-p^2_y},
\end{eqnarray}
where for main part of the Fermi surface of the Q2D electrons (3)
\begin{equation}
p^0_x(p_y) \sim p_F.
\end{equation}
Eq.(8) allows us to use quasi-classical approximation for the
electron Hamiltonian (6) and electron wave function (7):
\begin{eqnarray}
&&\frac{1}{2m} \biggl[ \frac{p^2_F - p^2_y}{2m} \pm 2 i p^0_x(p_y)
\frac{d}{dx} + \frac{p^2_y}{2m} - 2 t_{\perp} \cos \biggl(p_z c^*
\nonumber\\
&&- \frac{\omega_c}{v_F} x \biggl) \biggl] \
\Phi_{\epsilon}^{\pm}(x,p_y,p_z) = (\epsilon + \epsilon_F)\
\Phi_{\epsilon}^{\pm}(x,p_y,p_z),
\end{eqnarray}
where energy $\epsilon$ is counted from the Fermi level. It is
easy to rewrite Eq.(9) in more convenient way:
\begin{eqnarray}
&&\biggl[ \pm i v^0_x(p_y) \frac{d}{dx} - 2 t_{\perp} \cos
\biggl(p_z c^* - \frac{\omega_c}{v_F} x \biggl) \biggl]
\Phi_{\epsilon}^{\pm}(x,p_y,p_z)
\nonumber\\
&&= \epsilon \ \Phi_{\epsilon}^{\pm}(x,p_y,p_z), \ \ \ v^0_x(p_y)
= p^0_x(p_y)/m .
\end{eqnarray}

As discussed above, we consider the case of relatively small
magnetic fields and high enough temperatures, where quantum
effects of electron motion between the conducting planes in a
magnetic field [8-10] are negligible. In this case, we can
consider in Eq.(10) only the first order terms with respect to the
magnetic field. As a result of this procedure, we obtain,
\begin{eqnarray}
&&-2t_{\perp} \cos \biggl(p_zc^* - \frac{\omega_c}{v_F}x \biggl)
\approx -2t_{\perp} \cos (p_zc^*)
\nonumber\\
&&-\biggl( \frac{2t_{\perp} \omega_c x}{v_F} \biggl) \sin
(p_zc^*),
\end{eqnarray}
and, therefore, Eq.(10) can be represented as
\begin{eqnarray}
&&\biggl[ \pm i v^0_x(p_y) \frac{d}{dx}  -2t_{\perp} \cos (p_zc^*)
-\biggl( \frac{2t_{\perp} \omega_c x}{v_F} \biggl) \sin (p_zc^*)
\nonumber\\
&&- \mu_B \sigma H \biggl] \
\Phi_{\epsilon}^{\pm}(x,p_y,p_z;\sigma) = \epsilon \
\Phi_{\epsilon}^{\pm}(x,p_y,p_z;\sigma),
\end{eqnarray}
where we take into account also the Pauli spin-splitting effects
in the field for spin up ($\sigma=+1$) and spin down
($\sigma=-1$), $\mu_B$ is the Bohr magneton. Eq.(12) can be
exactly solved:
\begin{eqnarray}
&&\Phi_{\epsilon}^{\pm}(x,p_y,p_z;\sigma) = \exp \biggl(\mp i
\frac{\epsilon x}{v^0_x} \bigg) \exp \biggl[\mp i \frac{2
t_{\perp}\cos (p_zc^*) x}{v^0_x} \biggl]
\nonumber\\
&&\times \exp \biggl(\mp \frac{\mu_B \sigma x}{v^0_x} \biggl) \exp
\biggl[ \mp i \frac{t_{\perp} \omega_c x^2}{v^0_x v_F} \sin(p_z
c^*) \biggl] \ .
\end{eqnarray}

For Hamiltonian (12), we have the following differential equations
to determine the electron Green's functions in the mixed
$(x,p_y,p_z)$ representation [11,12]:
\begin{eqnarray}
&&\biggl[i \omega_n  \mp i v^0_x(p_y) \frac{d}{dx}  +2t_{\perp}
\cos (p_zc^*) +\biggl( \frac{2t_{\perp} \omega_c x}{v_F} \biggl)
\sin (p_zc^*)
\nonumber\\
&&+ \mu_B \sigma H \biggl] g_{i \omega_n}^{\pm}(x,p_y,p_z, \sigma)
= \delta(x-x_1) \ .
\end{eqnarray}

In Eq.(14), $\omega_n$ is the so-called Matsubara frequency [12].
Let us solve Eq.(14) analytically. As a result, for the Green's
functions we obtain:
\begin{eqnarray}
&&g^{\pm}_{i \omega_n} ( x, x_1;p_y,p_z;\sigma) = - i \frac{ sgn(
\omega_n)}{v_x(p_y)}
 \exp \biggl[ \mp \frac{\omega_n (x-x_1)}{v_x(p_y)} \biggl]
\nonumber\\
&&\times \exp \biggl[\pm i \frac{2 t_{\perp}\cos (p_zc^*)
(x-x_1)}{v^0_x} \biggl]\exp \biggl[ \frac{\mp i \mu_B \sigma H
(x-x_1)}{v_x(p_y)} \biggl]
\nonumber\\
&&\times \exp \biggl[ \pm i \frac{t_{\perp} \omega_c
(x^2-x^2_1)}{v^0_x v_F} \sin(p_z c^*) \biggl] .
\end{eqnarray}

Let us derive the so-called linearized Gor'kov's equation [12] for
non-uniform superconductivity to determine superconducting
transition temperature as a function of a magnetic field,
$T_c(H)$. As a result, we obtain
\begin{eqnarray}
&&\Delta(x) =\frac{g}{2} \biggl< \int^{\infty}_{ |x-x_1| = d |\sin
\alpha|} \frac{2 \pi T dx_1}{v_F \sin \alpha \sinh \biggl( \frac{
2 \pi T  |x-x_1|}{ v_F \sin \alpha} \biggl) }
\nonumber\\
&&\times J_0 \biggl[ \frac{2 t_{\perp} \omega_c}{v^2_F \sin
\alpha} (x^2-x^2_1)\biggl]  \cos \biggl[ \frac{2  \mu_B H
(x-x_1)}{v_F \sin \alpha} \biggl]  \ \Delta(x_1) \biggl>_{\alpha}\
,
\end{eqnarray}
where $<...>_{\alpha}$ stands for averaging over angle $\alpha$,
$g$ is the effective electron-electron interactions constant, $d$
is the cut-off distance, $J_0(...)$ is the zero-order Bessel
function. To show that Eq.(16) does not contain singularity at
$\alpha=0$, below we introduce new variable of integration, $x_1=z
\sin \alpha +x$, and rewrite Eq.(16) in the following more
convenient way:
\begin{eqnarray}
&&\Delta(x) = g \biggl< \int^{\infty}_{d} \frac{2 \pi T d z}{v_F
\sinh \biggl( \frac{ 2 \pi T z}{ v_F} \biggl) } \cos \biggl[
\frac{2  \mu_B H z}{v_F} \biggl]
\nonumber\\
&&\times J_0 \biggl\{ \frac{2 t_{\perp} \omega_c}{v^2_F } [z(2x+z
\sin \alpha)]\biggl\}  \ \Delta(x +z \sin \alpha)
\biggl>_{\alpha}.
\end{eqnarray}

Let us first derive the GL slope of the parallel upper critical
magnetic field from Eq.(17). To this end, we expend the Bessel
function and the superconducting gap with respect to small parameter
$z \ll v_F/(\pi T_c)$:
\begin{eqnarray}
&&J_0 \biggl\{ \frac{2 t_{\perp} \omega_c}{v^2_F} [z (z \sin
\alpha +2x)] \biggl\} \approx 1-\frac{4 t^2_{\perp}
\omega^2_c}{v^4_F} x^2 z^2 \ ,
\nonumber\\
&&\Delta(x+z \sin \alpha) \approx \frac{1}{2} z^2 \sin^2 \alpha
\frac{d^2 \Delta(x)}{dx^2} \ .
\end{eqnarray}
After substituting (18) into integral of Eq.(17) and averaging over angle
$\alpha$ we obtain:
\begin{eqnarray}
&&\Delta(x) \biggl[ \frac{1}{g} - \int_d^{\infty}  \frac{2 \pi T d
z}{v_F \sinh \biggl( \frac{ 2 \pi T z}{ v_F} \biggl)} \biggl]
\nonumber\\
&&-\frac{1}{4} \frac{d^2 \Delta(x) }{dx^2} \int_0^{\infty} \frac{2
\pi T_c z^2 d z}{v_F \sinh \biggl( \frac{ 2 \pi T_c z}{ v_F}
\biggl)}
\nonumber\\
&&+ x^2 \Delta(x) \frac{4t^2_{\perp}\omega^2_c}{v^4_F}
\int_0^{\infty} \frac{2 \pi T_c z^2 d z}{v_F \sinh \biggl( \frac{
2 \pi T_c z}{ v_F} \biggl)} = 0 \ ,
\end{eqnarray}
where $T_c$ is superconducting transition temperature in the
absence of magnetic field, which satisfies the following equation:
\begin{equation}
 \frac{1}{g} = \int_d^{\infty}  \frac{2 \pi T_c d z}{v_F
\sinh \biggl( \frac{ 2 \pi T_c z}{ v_F} \biggl)} \ .
\end{equation}
Here, we also take into account that [13]:
\begin{equation}
\int^{\infty}_0 \frac{x^2 dx}{\sinh(x)}  = \frac{7}{3} \zeta(3) \ ,
\end{equation}
where $\zeta(x)$ is the Riemann zeta-function, and introduce
the parallel and perpendicular GL coherence lengths,
\begin{equation}
\xi_{\parallel} = \frac{\sqrt{7 \zeta(3)}v_F}{4 \sqrt{2} \pi T_c}, \ \ \
\xi_{\perp} = \frac{\sqrt{7 \zeta(3)} t_{\perp} c^*}{2 \sqrt{2} \pi T_c},
\end{equation}
correspondingly.
Now differential gap Eq.(19) can be rewritten as:
\begin{equation}
- \xi^2_{\parallel} \frac{d^2 \Delta(x)}{dx^2}
+ \biggl(\frac{2\pi H}{\phi_0}\biggl)^2 \xi^2_{\perp} x^2 \Delta(x)
-\tau \Delta(x) = 0,
\end{equation}
where $\phi_0 = \frac{\pi c}{e}$ is the magnetic flux quantum,
$\tau=\frac{T_c-T}{T_c}$. It is important that Eq.(23) can be
analytically solved [1] and expression for the GL upper critical
field slope can be analytically written:
\begin{equation}
H^{\parallel}_{c2} = \tau \biggl( \frac{\phi_0}{2 \pi
\xi_{\parallel} \xi_{\perp}} \biggl) = \tau \biggl[ \frac{8 \pi^2
c T^2_c}{7 \zeta(3) e v_F t_{\perp} c^*} \biggl] .
\end{equation}

Below, we consider the general Eq.(17) to determine the so-called superconducting
nucleus and the parallel upper critical magnetic field at zero temperature.
To this end, we rewrite Eq.(17) for $T=0$:
\begin{eqnarray}
&&\Delta(x) = g \biggl< \int^{\infty}_{d} \ \frac{d z}{z} \
 J_0 \biggl\{ \frac{2 t_{\perp} \omega_c}{v^2_F } [z(2x+z
\sin \alpha)]\biggl\}
\nonumber\\
&&\times   \cos \biggl[
\frac{2  \mu_B H z}{v_F} \biggl] \ \Delta(x +z \sin \alpha)
\biggl>_{\alpha}.
\end{eqnarray}
Note that Eq.(25) is rather general, since it contains not only
orbital destructive effects against superconductivity in a
magnetic fields but also the spin-splitting Pauli effects against
singlet $s$-wave superconductivity. In this Letter, we are
interested only in the orbital effects and will disregard the
spin-splitting ones. In this case, it is convenient to introduce
the following new variables,
\begin{equation}
\tilde z = \frac{\sqrt{2t_{\perp} \omega_c}}{v_F} z , \ \ \ \tilde
x = \frac{\sqrt{2t_{\perp} \omega_c}}{v_F} x,
\end{equation}
and rewrite Eq.(25) using new variables as
\begin{eqnarray}
&&\Delta(\tilde x) = g \biggl< \int^{\infty}_{\frac{
\sqrt{2t_{\perp}\omega_c}}{v_F}d}
\ \frac{d \tilde z}{\tilde z} \
 J_0 [\tilde z(2 \tilde x + \tilde z
\sin \alpha)]
\nonumber\\
&&\times\Delta(\tilde x + \tilde z \sin \alpha)
\biggl>_{\alpha}.
\end{eqnarray}
We stress that solution of Eq.(27) (i.e., the so-called
superconducting nucleus [1,2]) corresponds to the parallel upper
critical magnetic field at zero temperature. Numerical solution of
Eq.(27) (see Fig.1) gives the following result for the parallel
upper magnetic critical fields in terms of the GL slope (24):
\begin{equation}
H^{\parallel}_{c2}(0) \approx 0.75 \biggl[ \frac{8 \pi^2 c
T^2_c}{7 \zeta(3) e v_F t_{\perp} c^*} \biggl] = 0.75 \ | d
H_{c2}^{\parallel}/ d T|_{T_c} T_c \ .
\end{equation}

\begin{figure}[t]
\centering
\includegraphics[width=0.5\textwidth]{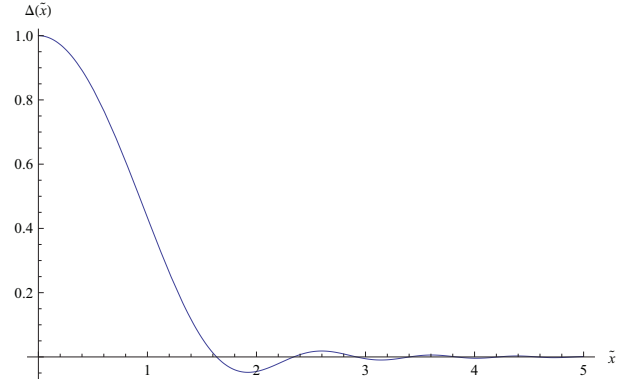}
\caption{Solution of Eq.(27) for the Q2D conductor (3) in the
parallel magnetic field (4) is shown. We pay attention to the fact
that the solution is not of the Gaussian form, moreover it changes
its sign several times with changing variable $\tilde x$.}
\end{figure}

Here, we consider the perpendicular upper critical magnetic field
of the Q2D superconductor (3). Therefore, in this case, we choose
magnetic field and vector potential in the the following form:
\begin{equation}
{\bf H} = (0,0,H), \ \ \ {\bf H} = (0,Hx,0).
\end{equation}
Using exactly the same steps and procedures as before for the
parallel field, it is possible to obtain for the perpendicular
field the following linearized Gor'kov's equation [2,12] for
non-uniform superconductivity:
\begin{eqnarray}
&&\Delta(x) = g \biggl< \int^{\infty}_{d} \frac{2 \pi T d z}{v_F
\sinh \biggl( \frac{ 2 \pi T z}{ v_F} \biggl) } \cos \biggl[
\frac{2  \mu_B H z}{v_F} \biggl]
\nonumber\\
&&\times \cos \biggl\{ \frac{eH}{c} [z\sin \alpha(2x+z
\cos \alpha)]\biggl\}  \ \Delta(x +z \cos \alpha)
\biggl>_{\alpha}.
\end{eqnarray}

Then, by means of the same method, as for the parallel magnetic
field described in detail above, we obtain the similar GL equation
in the perpendicular magnetic field (29):
\begin{equation}
- \xi^2_{\parallel} \frac{d^2 \Delta(x)}{dx^2} + \biggl(\frac{2\pi
H}{\phi_0}\biggl)^2 \xi^2_{\parallel} x^2 \Delta(x) -\tau
\Delta(x) = 0.
\end{equation}
Analytic solution of the Eq.(31) results in the following
formula for the GL slope of the perpendicular upper critical
magnetic field:
\begin{equation}
H^{\perp}_{c2} = \tau \biggl( \frac{\phi_0}{2 \pi
\xi^2_{\parallel}} \biggl) = \tau \biggl[ \frac{16 \pi^2 c
T^2_c}{7 \zeta(3) e v_F^2} \biggl] .
\end{equation}
Repeating analogous analysis, as for the parallel magnetic field,
we can write equation to determine the perpendicular upper
critical magnetic field in the form:
\begin{eqnarray}
&&\Delta(x) = g \biggl< \int^{\infty}_{d} \ \frac{d z}{z} \
 \cos \biggl\{ \frac{eH}{c} [z \sin \alpha(2x+z
\cos \alpha)]\biggl\}
\nonumber\\
&&\times\Delta(x +z \cos \alpha) \biggl>_{\alpha},
\end{eqnarray}
which after introducing new variables,
\begin{equation}
\tilde z = \sqrt{\frac{eH}{c}} z , \ \ \ \tilde
x = \sqrt{\frac{eH}{c}} x,
\end{equation}
reduces to
\begin{eqnarray}
&&\Delta(\tilde x) = g \biggl< \int^{\infty}_{ \sqrt{\frac{
eH}{c}}d} \ \frac{d \tilde z}{\tilde z} \
 \cos [\tilde z \sin \alpha (2 \tilde x + \tilde z
\cos \alpha)]
\nonumber\\
&&\times\Delta(\tilde x + \tilde z \cos \alpha) \biggl>_{\alpha}.
\end{eqnarray}
It is possible to prove that $\Delta(\tilde x) = \exp(-\tilde
x^2)$ is the solution of Eq.(35) which gives the following value
of the perpendicular upper critical magnetic field:
\begin{equation}
H^{\perp}_{c2}(0) \approx 0.59 \biggl[ \frac{16 \pi^2 c  T^2_c}{7
\zeta(3) e v_F^2} \biggl] = 0.59  \ | d H_{c2}^{\perp}/ d T|_{T_c}
T_c \ .
\end{equation}

Let us discuss the possible applicability of the derived above
Eqs.(28) and (36), which predict an increase of the Q2D
anisotropy, $\gamma(T) =
\biggl[\frac{H^{\parallel}_{c2}(T)}{H^{\perp}_{c2}(T)}\biggl]$,
with decreasing temperature:
\begin{equation}
\lim_{T \rightarrow 0} \gamma(T) = \lim_{T \rightarrow 0} \biggl[
\frac{H^{\parallel}_{c2}(T)}{H^{\perp}_{c2}(T)} \biggl] = 1.27
\lim_{T \rightarrow T_c} \gamma(T) .
\end{equation}
Note that, first of all, in the Letter, we consider the case of a
clean Q2D superconductor, which is opposite to the so-called
Lawrence-Doniach model [14,15]. Therefore, in our case,
$\xi_{\perp} \gg c^*$ [see Eqs.(3) and (22)]. Secondly, the
calculated orbital effect is supposed to be stronger than the
Pauli spin-splitting effects [16-18] in a magnetic field and,
thus, there are no conditions for the appearance of the
Fulde-Ferrell-Larkin-Ovchinnikov phase [18-20]. We pay attention
that there exists important high-temperature superconductor,
$MgB_2$, where the above discussed conditions are fulfilled (see,
for example, Refs. [21,22]) and where the above mentioned increase
of $\gamma(T)$ is experimentally observed [5,6]. Our point of view
is that this phenomena in $MgB_2$ is partially due to the effect,
suggested in this Letter, and partially - due to the many-band
effects [6,23,24], suggested for its explanation earlier.

In conclusion, we stress that all calculations for a Q2D
superconductor in a magnetic field have been performed in the
framework of the Fermi liquid theory [1]. In Q2D case, it is
always possible to do, unlike Q1D one [25].

$^*$Also at: L.D. Landau Institute for Theoretical Physics, RAS, 2
Kosygina Street, Moscow 117334, Russia.

\end{document}